\input harvmac
\newcount\figno
\figno=0
\def\fig#1#2#3{
\par\begingroup\parindent=0pt\leftskip=1cm\rightskip=1cm
\parindent=0pt
\baselineskip=11pt
\global\advance\figno by 1
\midinsert
\epsfxsize=#3
\centerline{\epsfbox{#2}}
\vskip 12pt
{\bf Fig. \the\figno:} #1\par
\endinsert\endgroup\par
}
\def\figlabel#1{\xdef#1{\the\figno}}
\def\encadremath#1{\vbox{\hrule\hbox{\vrule\kern8pt\vbox{\kern8pt
\hbox{$\displaystyle #1$}\kern8pt}
\kern8pt\vrule}\hrule}}

\overfullrule=0pt

\Title{TIFR-TH/99-19; EFI-99-17}
{\vbox{\centerline{Holograms of Branes in the Bulk and}
\centerline{Acceleration Terms in SYM Effective Action}}}
\smallskip
\centerline{Sumit R. Das\foot{E-mail: das@theory.tifr.res.in}}
\smallskip
\centerline{\it Enrico Fermi Institute,}
\centerline{\it University of Chicago, Chicago, IL 60637, U.S.A.}
\smallskip
\centerline{and}
\smallskip
\centerline{\it Tata Institute of Fundamental Research
\foot{Permanent address}}
\centerline{\it Homi Bhabha Road, Mumbai 400 005, INDIA}

\bigskip

\medskip

\noindent

If the AdS/CFT correspondence is valid in the Coulomb branch, the
potential between waves on a pair of test branes in the bulk should be
reproduced by the relevant Yang-Mills theory effective action on the
boundary. Earlier work has provided evidence for this in the case of
constant gauge field brane waves. In this paper we provide concrete
evidence for an earlier proposal that the effects of exchange of
supergravity modes with nonzero momentum in the brane directions are
encoded in certain terms involving derivatives of the field strength
in the gauge theory effective action. We explicitly calculate the
force quadratic in the field strengths coming from the exchange of
non-zero momentum two form fields between two 3-branes in $AdS_5 \times
S^5$ to lowest nontrivial order in the momentum. We show that this is
exactly the same as that between the branes living in flat space. The
result is in agreement with the gauge theory effective action and
consistent with the non-renormalization property of this term. We
comment on the relationship of other ``acceleration'' terms in the SYM
effective action with quantities in supergravity.

\Date{April, 1999}

\def\adsf{{AdS_5 \times S^5}}
\def\roR{({r \over R})}
\def\Ror{({R \over r})}

\def\tF{{\tilde F}}
\def\g{\delta t}
\def\cN{\cal N}
\def\g{g_{YM}}
\def\fa{F_1}
\def\fb{F_2}
\def\tO{{\tilde O}}
\def\tT{{\tilde T}}
\def\tF{{\tilde F}}
\def\cJ{{\cal J}}
\def\ga{{G_1}}
\def\gb{{G_2}}
\def\fat{{\tilde F}_1}
\def\fbt{{\tilde F}_2}

\newsec{Introduction.}

The duality between Type IIB superstring theory on $\adsf$ and $\cN$ $=
4$ supersymmetric Yang-Mills theory in $3+1$ dimensions 
\ref\malda{J. Maldacena, Adv. Theor. Math. Phys. 2 (1998) 231,hep-th/9711200}
\ref\gkpw{S.S. Gubser, I.R. Klebanov and A.M. Polyakov, 
Phys. Lett. B428 (1998) 105, hep-th/9802109; E. Witten,
Adv. Theor. Math. Phys. 2 (1998) 253, hep-th/9802150.} 
is most commonly studied at the conformally
invariant point. Recently there has been some evidence that the
duality is valid in the Coulomb branch \foot{We will consider the gauge
theory to live on $R^{(3,1)}$ rather than $S^3 \times R$ so that there
is a Coulomb branch.}
as well \ref\dtay{M. Douglas and
W. Taylor, hep-th/9807225} \ref\das{S.R. Das, JHEP 9902 (1999) 012,
 hep-th/9901004.}
\ref\trivnew{A. Bilal and C. Chu, Nucl. Phys. B 547 (1999) 179, 
hep-th/9810195; 
P. Kraus, F. Larsen and S. Trivedi, JHEP 9903 (1999) 003, hep-th/9811120.}. 
A simple setting
for this is to study the force between waves on a pair of test 3-branes
in the bulk of $\adsf$ \dtay. On the supergravity side this force is due to
exchange of massless supergravity modes propagating in $\adsf$, while
on the Yang-Mills side this is read off from an effective action obtained
by intergrating out fluctuations around the relevant expectation values
of the scalars (representing the positions of the test branes) and
other fields (representing the waves on the branes). Consider for
simplicity brane waves made of gauge fields alone. Let $y^a$ denote
coordinates along the brane directions and $Z^i$ denote the coordinates
transverse to the branes. A pair of branes located at positions $Z_1^i$
and $Z_2^i$ is represented in the $U(N)$ Yang-Mills theory by the constant
expectation
value of the Higgs field $\phi^i$
\eqn\one{\phi^i = \pmatrix{z_1^i & 0 & 0 \cr 0 & z_2^i & 0 \cr
0 & 0 & {\bf 0}_{(N-2)\times (N-2)}}}
where the expectation values $z^i$ are related to the physical locations 
$Z^i$ by
\eqn\ooone{z^i = Z^i~l_s^{-2}}
where $l_s$ is the string length.
The gauge field excitations $(\fa)_{ab}$ and $(\fb)_{ab}$ on the two 
branes are
represented by a $U(N)$ gauge field
\eqn\two{F_{ab}(y)  = \pmatrix{(\fa)_{ab} (y)  & 0 & 0 \cr 0 & 
(\fb)_{ab} (y)  & 0 \cr
0 & 0 & {\bf 0}_{(N-2)\times (N-2)}}}
When $N=2$ this represents branes in flat space, while for large $N$ and
in the Maldacena scaling limit these are branes in $\adsf$.

When the gauge fields on the branes are constant, the terms in the 
one-loop effective
action in the gauge theory involving cross-terms in $F_1$ and $F_2$ reads
\eqn\three{{\g^4\over \rho^4}[O^\phi_1 O^\phi_2 + O^\chi_1 O^\chi_2 + 
2 T^{ab}_1 T_{2~ab} + 2 O^{ab}_1 O_{2~ab}|_4]}
where $\g$ is the Yang-Mills coupling constant and for each $i = 1,2$ 
\eqn\four{\eqalign{& O^\phi_i = {1\over 4}(F_i)^{ab}(F_i)_{ab}\cr
& O^\chi_i = {1\over 8}\epsilon_{abcd} (F_i)^{ab}(F_i)^{cd}\cr
& T^{ab}_i = {1\over 2}[(F_i)^a_c (F_i)^{cb}
- {1\over 2} \eta^{ab} (F_i)^{cd}(F_i)_{cd}]\cr
& O^{ab}_i = {1\over 2}[(F_i)^{ab} + (F_i)^{ac}
(F_i)_{cd}(F_i)^{db} - {1\over 4} (F_i)^{ab}
(F_i)^{cd}(F_i)_{cd}]}}
while
\eqn\five{ \rho^2 = \sum_{i=1}^6 (z^i_1 - z^i_2)^2}
In the above expressions we have normalized the fields so that the kinetic term
in the gauge theory action does not contain any power of $\g$.
In \three\ the subscript $|_4$ means the term with four factors of the
gauge field. 

Note that the one loop result is identical for $SU(2)$ and $SU(N)$ since
the particles which run around the loop must be charged under both the
$U(1)$'s in which $z_1,z_2,F_1,F_2$ lie. 
For $SU(2)$ this one loop result is known to be exact
\ref\dine{M. Dine and N. Seiberg, Phys. Lett. B409 (1997) 239,
hep-th/9705057;
S. Paban, S. Sethi and M. Stern, Nucl. Phys. B534 (1998) 137,
hepth/9805018 and JHEP 06 (1998) 012, hep-th/9806028.}. For $SU(N)$ this
is also exact at a generic point on the Coulomb branch when one includes
a sum over the various $U(1)$ factors
\ref\lowe{D. Lowe, JHEP 9811 (1998) 009, hep-th/9810075; 
D. Lowe and R. von Unge, JHEP 9811 (1998) 014,
hep-th/9811017}. For our situation, in which
$SU(N) \rightarrow SU(N-2) \times [U(1)]^2$ these nonrenormalization theorems
are still expected to be valid \dtay. This means that the result 
\three\ is valid at strong coupling
where a comparison with supergravity can be made.

For $SU(2)$ the various terms in \three\ have a straightforward
supergravity interpretation in terms of exchange of supergravity
modes with zero momentum in the brane directions. Note that $\rho$ is
related to the
transverse distance $\Delta Z$ between the branes by 
\eqn\onea{\Delta Z = \rho l_s^2}
while the 10d gravitational coupling
constant $\kappa$ is related to the gauge coupling by (upto a numerical
constant)
\eqn\twoa{\kappa = \g^2 l_s^4} 
Thus 
\eqn\threea{ {\g^4 \over \rho^4} = {\kappa^2 \over (\Delta Z)^4}} 
However $1/(\Delta Z)^4$ 
is the static Coulomb propagator in the flat six dimensional
transverse space, while the operators $O^\phi, O^\chi, T^{ab}$ and
$O^{ab}$ are related to the operators ${\tilde O}$ in the brane action 
which couple to the
dilaton, RR scalar, longitudinally polarized (ten dimensional)
graviton and longitudinally polarized 2-form fields respectively by
\eqn\foura{\eqalign{& \tO^\phi = \kappa O^\phi~~~~~~\tO^\chi = \kappa O^\chi
~~~~~~~\tT^{ab} = \kappa T^{ab} \cr
&(\tO^{ab}_i)^{flat} = {{\sqrt{\kappa}}\over 2}(F_i)^{ab} + 
{\kappa^{3/2} \over 2}[(F_i)^{ac}
(F_i)_{cd}(F_i)^{db} - {1 \over 4} (F_i)^{ab}
(F_i)^{cd}(F_i)_{cd}]}}
where the supergravity fields have been normalized such that there is
no power of $\kappa$ in their kinetic terms. Using the above formulae
it may be checked  that the gauge theory answer for the $O(F^4)$ terms 
is exactly of the same form 
as the supergravity answer. This is similar to the way velocity dependent
forces between branes are reproduced from the relevant gauge theory
and has played an important role in M(atrix) theory \ref\matrix{
M. Douglas, D. Kabat, P. Pouloit and S. Shenker, Nucl. Phys.
B 485 (1997) 85 (hep-th/9608024);  T. Banks,
W. Fischler, S. Shenker and L. Susskind, 
Phys. Rev. D55 (1997) 5112, hep-th/9610043; 
M. Douglas, J. Polchinski and
A. Strominger, JHEP 12 (1997) 003 (hep-th/9703031);
K. Becker and M. Becker,
Nucl. Phys. B 506 (1997) 48; K. Becker, M. Becker, J. Polchinski and
A. Tseytlin, Phys. Rev D 56 (1997) 3174, hep-th/9706072.; 
I. Chepelev and A. Tseytlin, Nucl. Phys. B515 (1998) 73, hep-th/9709087;
D. Kabat and W. Taylor, Phys. Lett. B426 (1998) 297, hep-th/9712185; 
W. Taylor
and M. Raamsdonk, hep-th/9812239 and hep-th/9904095.}
In particular, 
the factors of $\g$ and $\rho$ in the former
give the correct factors of $\kappa$ and $\Delta z$ in the latter.

For $SU(N)$ and in the scaling limit of \malda, 
the result is due to exchange of supergravity modes
propagating in $\adsf$, but we expect that the final answer is the
same as that in flat space ! In \dtay\ it was shown that the zero
(brane) momentum propagators for the dilaton, RR scalar and the
longitudinal graviton \foot{These are all minimally coupled scalars in
the transverse space} in $\adsf$ are the same as those in flat
space. Their couplings to the brane fields are also identical in
$\adsf$ and flat space, which is related to the fact that they couple
to dimension four operators. This explains why the first three terms
in
\three\ come from supergravity exchange in $\adsf$.

In contrast, the propagators of the 2-form fields in $\adsf$
are quite different from those in flat space \das, even at zero
momentum. In particular the $NS-NS$ and $R-R$ 2-forms mix with each
other in the $\adsf$ background via the five form background
field. 

Let us use the standard form of the $\adsf$ metric
\eqn\six{ds^2 = \roR^2  [dy \cdot dy] + \Ror^2
[dr^2 + r^2 \sum_{i=1}^5 f_i(\theta_i) (d\theta_i)^2]}
We will use the following conventions. The ten dimensional
coordinates will be denoted by $y^\mu, \mu=0,\cdots 9$. Out of
these we denote the brane worldvolume directions
by $y^a, a = 0,\cdots 3$.
The remaining six transverse coordinates
$y^5 \cdots y^9$ will be relabelled as $Z^i, i = 1,\cdots 6$.
$r = {\sqrt{\sum_{i=1}^6 (Z^i)^2}}$ is the
radial coordinate in the transverse space and $\theta_i$ are
angles on the $S^5$. $(r, \theta_i)$ are related to the cartesian
coordinates $Z^i$ in the transverse space by the standard
transformations and the metric coefficients $f_i (\theta_i)$ are
determined from these transformations. 

We combine the two 2-form
fields into a complex field $B_{\mu\nu}$
\eqn\seven{B_{\mu\nu} = B^{NSNS}_{\mu\nu} + i B^{RR}_{\mu\nu}}
Then the propagator with zero momentum along $y^a$ is given
by \das\
\eqn\eight{\eqalign{G^{(0)}_{ab,cd}(Z_1,Z_2) & = \int d^4 y <B^*_{ab}(y,Z_1)
B_{cd} (y, Z_2)> \cr
& = {1\over 8\pi^3 |Z_1 - Z_2|^4}[a(Z_1,Z_2)(\eta_{ac}
\eta_{bd} - \eta_{ad}\eta_{bc}) + b(Z_1,Z_2) \epsilon_{abcd}]}}
where
\eqn\nine{\eqalign{& a(Z_1,Z_2) = ({r_1 \over R})^4 + ({r_2 \over R})^4 \cr
& b(Z_1,Z_2) = i [({r_1 \over R})^4 - ({r_2 \over R})^4]}}
Note that the propagator depends on the individual brane locations.
In flat space one has
\eqn\ten{G^{(0)}_{ab,cd}(Z_1,Z_2) = {1\over 4\pi^3 |Z_1 - Z_2|^4}[\eta_{ac}
\eta_{bd} - \eta_{ad}\eta_{bc}]}

In a similar way, the couplings of these modes to the brane fields are also
different in $\adsf$ \ref\dtrivedi{S.R. Das and S.P. Trivedi,
Phys. Lett. B 445 (1998) 142, hep-th/9804149} - 
this is related to the fact that these couple to
dimension six operators. The operator $(\tO^{ab}_i)^{flat}$ is now
modified to 
\eqn\sixa{(\tO^{ab}_i)^{Ads} = {{\sqrt{\kappa}}\over 2}(F_i)^{ab} + 
{\kappa^{3/2}\over 2}({R \over r})^4[(F_i)^{ac}
(F_i)_{cd}(F_i)^{db} - {1 \over 4} (F_i)^{ab}
(F_i)^{cd}(F_i)_{cd}]}
This also depends on the individual brane locations.

Nevertheless it was shown in \das\ that the
final result for the interaction potential between the branes 
with constant gauge fields on them is
identical to that in flat space and hence reproduced precisely by the
Yang-Mills effective action. This happens due to cancellations which
are not quite understood. In fact, even in flat space, the operator
$O^{ab}$ which couples to the 2-form field contains a term linear in
the field strength and one might expect a term which is quadratic in
$F_{ab}$.  From the gauge theory point of view such a term is not
present - a reflection of the nonrenormalization of the coupling
constant. In the supergravity calculation this again comes about due
to a cancellation between the NS-NS and R-R exchanges and this
persists in $\adsf$.  These results provide rather non-trivial
evidence for validity of the AdS/CFT correspondence in the Coulomb
branch.

When the waves on the branes are not constant, supergravity modes
which are exchanged will carry nonzero momentum in the brane
directions and the result will be sensitive to causal propagation in
the bulk.  However two branes situated at the same point in $S^5$ and
separated along the ``radial'' direction in $AdS_5$ correspond to the
same location in the boundary gauge theory and one might wonder how
the boundary theory knows about causal propagation in the bulk.  

In \das\ it was proposed that some of these effects are encoded in certain
terms in the Yang-Mills effective action which involve derivatives of
the gauge fields. We will refer to such terms as ``acceleration
terms'' since similar terms with the gauge fields replaced by the
Higgs field represent accelerations of the probe branes in the
transverse space. In this context such terms were studied in
\ref\peri{V. Periwal and R. von Unge, Phys. Lett B430 (1998) 71,
hep-th/9801121.}.
The simplest such term involving only gauge fields in fact comes
from the exchange of the 2-form field. For waves on branes situated in
flat space which are dependent only on time (and not the other brane
coordinates) it was shown in \das\ that supergravity predicts a term
proportional to $(\partial F)^2$ in the effective action. This comes
from the expansion of the (brane) momentum space propagator of the
2-form field around the static propagator.  Such a term is indeed
present in the same multiplet as the $F^4$ term \ref\roceka{B. de Wit,
M.T. Grisaru and M. Rocek, Phys. Lett. B374 (1996) 297,
hep-th/9601115; U. Lindstrom, F. Gonzalez-Rey, M. Rocek and R. von
Unge, hep-th/9607089; F. Gonzalez-Rey and M. Rocek, Phys. Lett. B434
(1998) 303, hep-th/9804010;
I.L. Buchbinder and S.M. Kuzenko, Mod.Phys.Lett. A13 (1998) 1623.}
\ref\rocekb{F. Gonzalez-Rey, B. Kulik, I.Y. Park
and M. Rocek, Nucl. Phys. B 544 (1999) 218, hep-th/9810152.} 
and is also one-loop exact, which
makes a comparison with supergravity meaningful.

While these results are indicative, they do not establish the proposal
even for this term. First, with the brane waves dependent on time
alone, the detailed Lorentz structure of the term is invisible while
the gauge theory predicts a specific Lorentz structure. Secondly the
above calculation was performed in flat space - i.e. with no other
brane present - where the relevant gauge
theory is $SU(2)$. The nonrenormalization
of this specific one loop term in the gauge theory predicts that the
same answer should follow in $\adsf$ and it is not clear whether the
cancellations which made the $F^4$ terms work are operative in this
context.

In this paper we provide further evidence in favor of this proposal.
We perform an explicit calculation of the $(\partial F)^2$ term in the
supergravity potential between branes in $\adsf$. We show, again due
to remarkable cancellations, that the result is the same as that
between branes in flat space and in exact agreement with the
predictions of Yang-Mills theory. This particular term vanishes when
the gauge fields are on-shell (i.e. $\partial_\mu F_i^{\mu\nu} = 0$),
but is present in an off-shell $N=2$ supersymmetric effective action
\rocekb.  As we shall see, the supergravity calculation requires that
the {\it total} current which couples to the 2-form field is
divergenceless.  This means that this particular $(\partial F)^2$ term
cancels with terms higher order in $F$ which come from other terms in
the divergenceless current.  Thus in SYM effective action there must
be cancellations between terms with different powers of the gauge
fields !  We then examine other possible derivative terms (which
generally survive on-shell) in the Yang-Mills effective action and
comment whether these terms should follow from supergravity.

\newsec{The strategy}

Consider the contribution to the force between the two probe 3-branes
coming from the exchange of a single supergravity mode.
Symbolically the interaction action is given by 
\eqn\bone{S_{eff}^{sugra}  = 
\int d^4 y \int d^4 y'~~[J^I (y,Z_1) \Delta_{IJ}(y-y',Z_1,Z_2)
J^J(y',Z_2) + (Z_1 \rightarrow Z_2)]}
Here $J^I(y,Z_1)$ denotes the current coupling to the supergravity mode
labelled by $I$ on the brane located at $Z_1$ and similarly for $J^J(y',Z_2)$.
$ \Delta_{IJ}(y-y',Z_1,Z_2)$ denotes the (retarded) propagator matrix of the
supergravity modes between the points $(y,Z_1)$ on the first brane
and $(y',Z_2)$ on the second brane. In writing down \bone\ we have used
translation invariance along the brane directions $y^a$.
With only gauge fields excited, these
currents are some expressions made out of these gauge fields. We want to
see whether $S_{eff}^{sugra}$ is 
reproducible in a gauge theory calculation. 

The expression $S_{eff}^{sugra}$ is of course 
an integral of a bi-local quantity, whereas
the gauge theory effective action is a sum of local terms arranged
usually as a derivative expansion. To make a connection between the
two it is easiest to go to momentum space on the brane but remain in
the transverse position space. We thus write
\eqn\bbone{\eqalign{&J^I(y,Z) = \int d^4p~e^{-ip\cdot y}J^I(p,Z)\cr
&\Delta_{IJ} (y-y',Z,Z') = \int d^4p~e^{-ip \cdot (y-y')}
~\Delta_{IJ} (p,Z,Z')}} 
Then \bone\ becomes
\eqn\bbtwo{S = \int d^4p~(J^I)^*(p,Z_1)\Delta_{IJ}(p,Z_1,Z_2)J^J(p,Z_2) + 
(Z_1 \rightarrow Z_2)}
We can now expand the propagator in powers of momentum and get an expression
which (after transforming back to position space on the brane) becomes
an expansion in derivatives of the currents.
\eqn\bthree{\eqalign{S = \int d^4 y_0 [& J^I (y_0, Z_1) \Delta^{(0)}_{IJ} 
(Z_1,Z_2) J^J(y_0, Z_2) \cr
& + {1\over 2} (\partial_{y^a} J^I (y_0, Z_1))(\partial_{y^b} J^J (y_0, Z_2))
\Delta^{ab}_{IJ} (Z_1, Z_2) + \cdots]}}
where 
\eqn\bfour{ \Delta^{(0)}_{IJ} (Z_1, Z_2) = \int d^4 s \Delta_{IJ} (s;Z_1,Z_2)}
is the zero momentum propagator and 
\eqn\bseven{\Delta^{ab}_{IJ}(Z_1,Z_2) = -[{\partial^2
\over \partial p_a \partial p_b} \Delta_{IJ} (p;Z_1,Z_2)]_{p = 0}}
The ellipses in \bthree\ denote terms with higher derivatives on the currents.

For such a supergravity calculation to be reproducible in gauge theory
one must be able to calculate at strong couplings in the gauge theory.
At present the only possibility is to concentrate on terms in the
effective action which are not renormalized, so that a perturbative
calculation in the gauge theory actually yield the strong coupling
answers. Indeed there is such a term in the $N=2$ off-shell effective
action which is quadratic in the field strength \rocekb\
\eqn\beight{ \int d^4 y {1\over \rho^2}[(\partial_a F^{ab})(\partial^c F_{cb})
+ (\partial_a \tF^{ab})(\partial^c \tF_{cb})]}
where $\rho$ is defined in \five\ and $F_{ab} = (\fa)_{ab} - (\fb)_{ab}$
while
\eqn\bnine{ \tF_{ab} = {1\over 2}\epsilon_{abcd}F^{cd}}
This term is in the same multiplet as the $F^4/\rho^4$ term and is also
one-loop exact. As a result this is identical for $SU(2)$ and $SU(N)$
gauge theories. Therefore it should follow from supergravity calculation
of the force between probe branes both in $\adsf$ and in flat space. In
the next section we will show that this is indeed true.

\newsec{Acceleration term from 2-form exchange}

Since the only supergravity mode which couples linearly to the gauge
fields on the brane is the 2-form field, we need to evaluate terms like
\bthree\ where the indices $(I,J)$ now refer to those of the 2-form.
As already shown in \das, the term quadratic in the field strengths
and containing no derivatives vanishes due to cancellations between
the NS-NS and RR 2-form exchanges. We are interested in the second
term of \bthree.

The piece of the 2-form propagator $<B^*_{ab}(p;Z_1)B_{gh}(p,Z_2)>$
which is quadratic in the brane
momenta, which we denote by $G^{(1)}_{ab,gh} (p;Z_1,Z_2)$,
is calculated in Appendix A with the final result
\eqn\cten{ G^{(1)}_{ab,gh} (p;Z_1,Z_2)
= \int d^6Z~({R \over r})^8~{1\over |Z-Z_1|^4}{1\over |Z_2-Z|^4}M_{ab,gh}
(p;Z,Z_1,Z_2)}
where 
\eqn\celeven{\eqalign{M_{ab,gh}(p;Z,Z_1,Z_2)
= & 2[p^2 a_1 a_2 (\eta_{ag} \eta_{bh} - \eta_{ah}\eta_{bg})\cr
& + (a_1a_2 + b_1b_2)(p_ap_h \eta_{bg}-p_ap_g \eta_{bh}+
p_bp_g \eta_{ah} - p_bp_h\eta_{ag})\cr
& + (b_2 a_1+b_1 a_2) p^2 \epsilon_{abgh} \cr
& + a_1b_2 (p_a p^f \epsilon_{bfgh} - p_bp^f \epsilon_{afgh})\cr
& + a_2 b_1(p_hp^c \epsilon_{abcg} - p_g p^c \epsilon_{abch})]}}
where we have used a shorthand notation :
\eqn\ctwelve{\eqalign{&a_1 = a(Z_1,Z)={1\over R^4}(r^4 + r_1^4)~~~~~~
a_2= a(Z,Z_2) = {1\over R^4}(r^4 + r_2^4) \cr
&b_1 = b(Z_1,Z) = -i{1\over R^4}(r^4 - r_1^4)~~~~~~b_2= b(Z,Z_2)
= -i{1\over R^4}(r_2^4 - r^4)}} 
This piece of the propagator, like the zero momentum piece, depends
on the individual locations of the two branes in a rather complicated
way.

The coupling of the 2-form field to the probe three-branes may be
read off from the DBI-WZ action of a single three brane in the $\adsf$
background. This coupling term is \dtrivedi\
\eqn\done{{\sqrt{\kappa}}
\int d^4 y \int d^6 Z ~[B^{ab} (y,Z) \cJ^*_{ab} (y,Z) + c.c.]}
where, in brane momentum space
\eqn\dtwo{\eqalign{\cJ^{ab}(p) =& [(\fa)^{ab}(p) + \kappa
({R^4 \over r_1^4})\ga^{ab}(p) 
+ i(\fat)^{ab} (p) ] \delta^6 (Z-Z_1) \cr
& + [(\fb)^{ab} (p)  + \kappa ({R^4 \over r_2^4})\gb^{ab} (p)  + 
i(\fbt)^{ab} (p) ]
\delta^6 (Z-Z_2) + \cdots}}
and we have defined the quantity
\eqn\dthree{G_i^{ab} = (F_i)^{ac} (F_i)_{cd}(F_i)^{db} 
- {1 \over 4} (F_i)^{ab}(F_i)^{cd}(F_i)_{cd}}
and the ellipses denote terms which are $O(F^4)$ and higher.  We have
to solve the classical equations of motion in the presence of this
current using the above propagator.  Consistency requires that
$\partial_a \cJ^{ab} = 0$.  It is easy to check that the equations of
motion for the mixed components of the 3-form field field strength are
obeyed by $B_{ij} = B_{ai} = 0$ (as has been assumed) and the solution
for $B_{ab}$ obtained from the above propagator, provided this
condition is satisfied.

Since we are interested in the term which has the structure 
$(\partial F_1)(\partial F_2)$, we can ignore the $G^{ab}$ piece in the
current. Then the full interaction action, upto terms with two derivatives 
on the gauge fields is obtained by combining \done,\dtwo,\bthree
and using the expressions \eight,\bseven\ and \cten\
\eqn\dfour{\eqalign{S_{eff}^{sugra} = \kappa
\int d^4 p [(\fa)^{ab}(p)-i(\fat)^{ab}(p)]& [G^{(0)}_{ab,gh}(Z_1,Z_2)
+ G^{(1)}_{ab,gh}(p;Z_1,Z_2)]\cr
& [(\fb)^{gh}(p) + i (\fbt)^{gh}(p)]
+ c.c.}}
It is easy to check that the term in \dfour\ involving $G^{(0)}$
vanishes, as already noted
in \das. The calculation of the second term is outlined in Appendix B.

Remarkably, the dependence on individual brane locations neatly cancel
in the final answer and we get
\eqn\dten{S_{eff}^{sugra} = {\kappa\over |Z_1 - Z_2|^2}\int d^4 y
[(\partial_a F^{ab})(\partial^c F_{cb})
+ (\partial_a \tF^{ab})(\partial^c \tF_{cb})]}
Using the relationship between the gauge and Yang-Mills couplings and
that between the physical distance and the Higgs values given in \onea\
and \twoa\ the supergravity expression is exactly the same, upto a numerical
factor which we did not calculate, as the gauge theory effective action
\beight.

The result \dten\ for branes separated in $\adsf$ is in fact identical
to the flat space answer. This may be seen by simply noting that in
flat space we should put $a_1 = a_2 = 2$ and $b_1 = b_2 = 0$ in the
above formulae, as seen from a comparison of the zero momentum
propagators \eight\ and \nine.  On the other hand there is no factor
of $({R \over r})^8$ in the term $S_1$ of the action. Thus we again
get back precisely the same final expression \dten. Even though we
have not been careful about numerical factors, it is important to note
that the flat space answer is {\it exactly} the same as the $\adsf$
answer. We have not bothered to calculate an overall numerical factor
in both.

\newsec{Other acceleration terms}

Once again remarkable cancellations conspired to yield a supergravity
derivation of a specific acceleration term of the gauge theory effective
action, providing an yet non-trivial evidence for the validity of
AdS/CFT correspondence. There must be a simple reason behind these
cancellations, probably related to the supersymmetry of the theory.
However we have not been able to find that yet.

Do all such acceleration terms in the Yang-Mills effective action have a
supergravity origin ? Consider for example terms with two factors of
the gauge field. At one loop, a cursory look at the feynman diagram
shows that one would have an expansion of the form
\eqn\etwo{ \g^2 [F^2 {\rm log}(\rho) + \sum_{n=1}{(\partial^n F)^2
\over \rho^{2n}}]}
We know that the first term vanishes by supersymmetry and the term
with $n = 1$ is the one which we found to have a supergravity 
interpretation. In fact it would be disastrous if the first term
was non-zero. This is because while Yang-Mills theory knows about the
coupling $g$ and the Higgs value $\rho$, supergravity knows only about
the gravitational coupling $\kappa$ and the physical transverse distance
$(\Delta Z)$. These are related by the expressions \onea\ and \twoa.
Thus the only term in \etwo\ which can be entirely expressed in terms
of supergravity quantities is the term $n = 1$. When $\g, \rho$
is traded for $\kappa, \Delta Z$ the first term in \etwo\ would
have {\it negative} powers of the string length; if nonzero this
cannot have a sensible explanation. In a similar way the terms with
$n > 1$ will now contain positive powers of $l_s$ and could have 
a stringy origin. Indeed such terms would be renormalized.
By the same token, a possible two loop correction
to $(\partial F)^2/\rho^2$ would also have a negative power of
$l_s$ : indeed this is zero since the one loop contribution to this
term is exact. It appears that the nonrenormalization theorems
work precisely in a way which supports the gauge theory - supergravity
correspondence.

A term in the SYM effective action 
with $2m$ factors of $F$ and $2n$ derivatives at $l$ loops
would have a form (dictated by simple dimension counting) which may be
schematically written as
\eqn\ethree{ \g^{2m + 2(l-1)}~{(\partial^{n} F^{m})^2 \over f_{2(2m+n-2)}(z_1,
z_2)}}
In \ethree\ $f_{2(2m+n-2)}(z_1,
z_2)$ denotes a polynomial of
$z_1$ and $z_2$ (these are defined in \one) of weight $2(2m+n-2)$
where we have assigned weight one to $z_1$ and $z_2$. Note
that beyond one loop the answer will generally depend on $z_1$ and $z_2$
individually rather than just $\rho$. 
Converting this to $\kappa, l_s$ and $Z_1,Z_2$ we get
\eqn\efour{\kappa^{m + (l-1)}~l_s^{4(m+n-l-1)}{(\partial^{n} F^{m})^2 \over 
f_{2(2m+n-2)}(Z_1,Z_2)}}
Thus for $m+n < l+1$ we have negative powers of the string length, for
$m+n > l+1$ we have positive powers of string length while the answer
is expressible entirely in terms of $\kappa$ and $Z_1,Z_2$ only for
$m + n = l + 1$. It is tempting to speculate that the first class of
terms ($m + n < l + 1$) vanish while there would be nonrenormalization
theorems which ensure that a term with given $(m,n)$ will be exact at
the $(m+n-1)$-th loop level. Those with $m+n > l+1$ do not have
an interpretation in supergravity. In \dtay\ it was suggested that
such terms could be interpreted in terms of a stringy uncertainity 
principle.  
Examples of exact terms at one loop
are the ones dealt with in \dtay,\das, $F^4/\rho^4$ and the term 
discussed in this paper, $(\partial F)^2/\rho^2$. This also suggests
that a term with say four factors of the gauge fields and two derivatives
should appear at two loops. There are known renormalization theorems
for such ``$v^6$'' terms as well. In flat space supergravity they
cannot appear at the level of a single mode exchange. However, as we
will soon see such terms can originate due to single mode exchange
in $\adsf$.

One could similarly start from the supergravity side and ask whether
all the terms which appear as a result of expansion of the momentum space 
propagator can appear in the Yang-Mills theory. Let us consider, for
example, terms which appear from a single dilaton exchange. The zero 
momentum piece is one of the $F^4 / \rho^4$ terms considered above.
If the branes are placed in flat space, the leading correction to
this term along the lines of this paper will be of the form
\eqn\efive{\kappa^2 {(\partial F^2)^2 \over (\Delta Z)^2}}
It is easy to see that this {\it cannot} be expressed in terms of
Yang-Mills quantities.

The situation is quite different when the branes are placed in $\adsf$.
Recall that the zero momentum propagator for the dilaton in $\adsf$
is identical to that in flat space, viz
\eqn\esix{ G^{(0)}_\phi (p, Z_1,Z_2)  = {1 \over |Z_1 - Z_2|^4}}
However the nonzero momentum propagator is {\it not} the same as in flat
space. The free part of the action of the dilaton field in a
$\adsf$ background reads
\eqn\eseven{ \eqalign{ & S_\phi  = S_\phi^{(0)} + S_\phi^{(1)}\cr
& S_\phi^{(0)} = \int d^4 y ~d^6 Z \sum_{i=1}^6 (\partial_{Z^i} \phi)^2\cr
& S_\phi^{(1)} = \int d^4 y ~d^6 Z ({R \over r})^4 \sum_{a=1}^4 
(\partial_{y^a} \phi)^2}}
The nonzero brane momentum piece $S_\phi^{(1)}$ depends on the
location $r$. Treating $S_\phi^{(1)}$ as a perturbation and 
using the methods of this paper it is easy to see that
the leading correction to the dilaton propagator is given by
\eqn\eeight{ G^{(1)}_\phi (p,Z_1,Z_2) = 
p^2 \int d^6Z~({R \over r})^4 {1 \over |Z_1 - Z|^4}
{1 \over |Z - Z_2|^4}}
This leads to a contribution to the interaction action which is of
the form
\eqn\enine{S_{\phi~eff}^{sugra} = 
{\kappa^2 R^4 \over f_{6}(Z_1,Z_2)} \int d^4 y~
(\partial_a O_1^\phi)(\partial^a O_2^\phi)}
where $f_{6} (Z_1, Z_2)$ is a polynomial of weight $6$ (as defined
below equation (4.2)) and the
operator $O_i^\phi$ has been defined in \four. Now, the
$AdS$ scale $R$ is related to $\g$ and $l_s$ by
\eqn\eten{ R^4 = (\g^2 N) l_s^4}
Using this and the relations \onea\ and \twoa\ this expression may
be recast entirely in terms of Yang-Mills quantities
\eqn\eeleven{S_{\phi~eff} = 
{\g^6 N \over f_{6}(z_1,z_2)} \int d^4 y ~  
(\partial_a O_1^\phi)(\partial^a O_2^\phi)}
Such a term could come from a two loop diagram in the Yang-Mills
theory. It has an additional factor of $\g^2 N$ compared to a
one loop term (with four external legs) and therefore has the
expected behavior in the large-N expansion. It would be
interesting to see if the detailed structure of this term is
obtainable from gauge theory.

We have seen an example where only branes placed in $\adsf$ and not in
flat space seem to have a correspondence with Yang-Mills theory. This
is what one expects. At one loop it is difficult to see the difference
between $SU(N)$ and $SU(2)$ (i.e. between branes in $\adsf$ and branes
in flat space) because of a combination of the structure of the terms
and nonrenormalization theorems.  In a similar way other
acceleration terms are expected to come from higher loop processes in
gauge theory. We hope that the problems which plagued similar
comparisons in M(atrix) theory at higher loops \ref\draja{M. Dine and
A. Rajaraman, Phys. Lett.  B 425 (1998) 77; M. Fabbrichesi, G. Ferreti
and R. Iengo, JHEP 9806 (1998) 002; Y. Okawa and T. Yoneya,
Nucl. Phys.  B 538 (1998) 67 and hep-th/9808188; J. McCarthy,
L. Susskind and A. Wilkins, hep-th/9806136; R. Echols and J. Gray,
hep-th/9806109} will not appear in our case.

Such acceleration terms have been a subject of discussion in the
past. Note that the term $(\partial F)^2$ we have obtained vanishes on
shell. In fact, for similar terms involving Higgs fields, this point
has been discussed in detail in \peri. In spite of being zero on-shell,
this term is automatic in an off-shell and explicitly $N=2$ 
supersymmetric effective action.  In \rocekb\ it was shown that such
terms can be in fact removed by a suitable redefinition of the
fields. Even though our physical situation is different from that in
\peri\ and \rocekb\ it will be interesting to re-examine these issues
in the light of our demonstration that at least one such term can be
obtained from a supergravity calculation and reflects relativistic
propagators in the bulk.

The supergravity calculation, however, requires that the divergence of
the total current $\cJ^{ab}$ defined in \dtwo\ is zero. Extending our
calculation to include the $O(F^3)$ terms in the current
we find that there are
nonzero contributions with the Lorentz structure $(\partial_c G_{1,ab})
(\partial^c F_2^{ab})$ where the tensor $G_{ab}$ has been defined in
\dthree. Note that similar terms (with the derivatives contracted
among themselves) cancelled out in our $O(F^2)$ calculation (see
Appendix B, equation (8.2)). Furthermore the type of terms we
have obtained above now generalizes to the structures 
$(\partial_a \cJ^{*~ab}) 
(\partial^c \cJ_{cb})$ and $(\partial_a \cJ^{*~ab})
(\partial^c {\tilde{\cJ}}_{cb})$. By consistency of the solution,
these latter terms should vanish. Thus, at the end of the day supergravity
predicts that the $O(F^2)$ contribution we have calculated in fact
cancels with $O(F^4)$ terms. The acceleration terms which do survive from the
2-form exchange are higher order in field strengths. They come from
higher loop effects in the Yang-Mills theory and do not
vanish when the perturbative equations of motion are used. In
addition there would be acceleration terms coming from the exchange
of the other massless fields like the dilaton.

This is consistent with the fact that a field redefinition removes the
$(\partial F)^2$ term from the effective action.  However, this is a
nontrivial prediction about the Yang-Mills theory.  This would mean
that in the off shell SYM effective action there would be precise
relations between terms with different numbers of gauge fields and two
derivatives. Furthermore, the supergravity terms we are talking about
can be entirely expressed in terms of SYM quantities only when we are
in $\adsf$. This is because the factors of $({R \over r})^4$ in front
of the cubic pieces of the current are necessary to get the right
factors of $l_s$ required to reexpress the coefficient in terms of
$g_{YM}, z_1, z_2$. Thus these relations must be present for large-$N$
SYM, but not for $SU(2)$. 

\newsec{Comments}

In summary, we have provided quantitative evidence in favor of the
proposal that the effects of nonzero (brane) momentum exchange between branes
are encoded in a class of terms in the dual gauge theory effective action
involving derivatives of the gauge field strengths. 
Such terms should be expressible entirely in
terms of gravitational quantities and should therefore
appear only at a particular order of the loop expansion. Known 
non-renormalization theorems seem to be consistent with this idea.

The piece of physics which is not transparent in this treatment
is how gauge theory really encodes causal properties of the
bulk. Since there is an integration over the coordinates of each
brane one effectively gets a propagator which is the sum of
the retarded and advanced propagators. Furthermore we have
performed an expansion in powers of the brane momenta. It is difficult to
see the causal structure in such an expansion.
What we have seen, however is that
the gauge theory effective action knows that light
has a finite speed in the bulk - we havent been able to figure out
whether it moves forward or backward in time. However other recent
results on bulk causality \ref\causality{G. Horowitz and N. Itzhaki,
JHEP 9902 (1999) 010, 
hep-th/9901012; D. Kabat and G. Lifshytz, JHEP 9812 (1998) 002 (hep-th/9806214)
and hep-th/9902073;
D. Bak and S.J. Rey, hep-th/9902101.}
seem to show that the guage theory knows
the latter as well.

\newsec{Acknowledgements} I would like to thank M. Douglas,
A. Jevicki, D. Kabat, D. Lowe, S. Mathur, S. Sethi, W. Taylor and
S. Trivedi for discussions at various stages of this work and 
A. Tseytlin for a correspondence. I also
thank the Department of Physics at Brown University, Center for
Theoretical Physics, MIT and Enrico Fermi Institute, University
of Chicago for hospitality.

\newsec{Appendix A : The propagator}

The bulk action for the complex 2-form field in a $\adsf$ background is
\eqn\cone{S = \int d^{10}y~{\sqrt{g}}{1\over 12}[H^*_{\mu\nu\lambda}
H^{\mu\nu\lambda} + i F^{\alpha\beta\mu\nu\lambda}(H_{\mu\nu\lambda}
B_{\alpha\beta} -H^*_{\mu\nu\lambda} B_{\alpha\beta} - (c.c.))]}
where 
\eqn\ctwo{H_{\mu\nu\lambda} = \partial_\mu B_{\nu\lambda} +
\partial_\nu B_{\lambda\mu} + \partial_\lambda B_{\mu\nu}}
and $F^{\alpha\beta\mu\nu\lambda}$ is the self-dual 5-form field
strength. The background metric is given in \six\ and the background
5-form field is (in the same notation as above)
\eqn\cthree{ F^{rabcd} = {\epsilon^{abcd} \over R}({R \over r})^3}
We will be interested in field configurations in which only the longitudinal
components of $B_{\mu\nu}$ are nonzero : $B_{ai}=B_{ij} = 0$. Then the
action may be rewritten as
\eqn\cfour{S = S_0 + S_1}
where $S_0$ is the part of the action which depends only on derivatives
along the transverse direction, while $S_1$ is the part which depends
on longitudinal derivatives.
\eqn\cfive{\eqalign{& S_0 = \int d^{10}y~{\sqrt{g}}{1\over 4}
[(\partial_i B^*_{ab})
(\partial^i B^{ab}) + iF^{rabcd}((\partial_r B_{ab})B_{cd} - 
(\partial_r B^*_{ab})B_{cd} - (c.c.))] \cr
& S_1 = \int d^{10}y ~{\sqrt{g}}{1\over 4}(\partial_a B^*_{bc})
[(\partial^a B^{bc}) +(\partial^b B^{ca}) + (\partial^c B^{ab})]}}
$S_0$ is the part of the action which was used in \das\ to obtain the
zero brane momentum propagator given in \eight. Note that the
Chern-Simons term which mixes the NSNS and RR 2-forms is present only
in $S_0$. Using the explicit form of the metric and utilizing translation
invariance in the brane directions to perform a fourier transform into
(brane) momentum space we get for $S_1$
\eqn\csix{S_1 = {1\over 4}\int [d^4p][d^6 Z] ({R \over r})^8~
B^*_{ab}(p,Z) B_{ef}(p,Z)~E^{cdabef}~p_cp_d}
where
\eqn\cseven{B_{ab}(y^a,Z^i)= \int [{d^4 p \over (2\pi)^4}]~
B_{ab}(p,Z)~e^{-ip_ay^a}}
and the tensor $E^{cdabef}$ is defined as
\eqn\ceight{E^{cd,ab,ef} = \eta^{cd}\eta^{ae}\eta^{bf} +
\eta^{ad}\eta^{be}\eta^{cf} +  \eta^{bd}\eta^{ce}\eta^{af}}

The propagators for massless modes in $\adsf$ are quite involved
and several of them have been obtained recently \ref\freedman{
E. D'Hoker and D. Freedman, hep-th/9809179 and hep-th/9811257;
E. D' Hoker, D. Freedman, S.D. Mathur, A. Matusis and L. Rastelli,
hep-th/9902042.}.
However, as is clear from the discussion of the last section, in
order to compute the term in the interaction potential between 
branes which involves only two derivatives in the brane direction,
all we need is the Taylor expansion of the momentum space propagator.
In the above action, $S_0$ does not involve any power of the
brane direction momentum while $S_1$ involves two powers of the
momenta. Thus it is sufficient for us to treat $S_1$ as a perturbation
to $S_0$. The term we are interested in is in fact obtained by a
single insertion of $S_0$ into the zero momentum propagator \eight.

The piece of the propagator quadratic in the brane momenta is therefore
\eqn\cnine{G^{(1)}_{ab,gh} (p;Z_1,Z_2)
= p_{c'} p_{d'}\int d^6Z~({R \over r})^8~G^{(0)}_{ab,cd}(Z,Z_1)~
E^{c'd',cd,ef}~G^{(0)}_{ef,gh}(Z_2,Z)}
Using \eight\ and \ceight, \cten\ follows after a long and straightforward
manipulation.

\newsec{Appendix B : The interaction energy}

The second term in (3.7) evaluates to the following expression after a
long and straightforward calculation
\eqn\dfive{S_{eff}^{sugra} = 
\kappa\int d^6Z~({R \over r})^8 ~{1\over |Z-Z_1|^4}{1\over |Z_2-Z|^4}
\int d^4p~ {\cal F}(p;Z_1,Z_2)}
where
\eqn\dsix{\eqalign{{\cal F} = & 2 (2a_1a_2 + -i(b_2a_1+b_1a_2))(p^ap_a)
((\fa)^{ab}(\fb)_{ab} +(\fat)^{ab}(\fbt)_{ab})\cr
& - 8[(a_1a_2 + b_1 b_2)-i(a_1b_2-a_2b_1)][(p_a(\fa)^{ab})(p^c(\fb)_{cb})
+ (p_a(\fat)^{ab})(p^c(\fbt)_{cb})]}}
Note that we have defined ${\tilde F}_{ab} = {1\over 2}\epsilon_{abcd}F^{cd}$
so that $(\fa)^{ab}(\fb)_{ab} = - (\fat)^{ab}(\fbt)_{ab}$. Thus the
first line in \dsix, which is proportional to $p^2$, 
vansihes. The second term has a coefficient which
becomes, after using the definitions of $a_1,a_2,b_1,b_2$ in \ctwelve\
and \nine\ \eqn\dseven{\eqalign{&(a_1a_2 + b_1 b_2) = {2 \over
R^8}(r^8 + r_1^4 r_2^4)\cr & i(a_1b_2-a_2b_1) = -{2 \over R^8}(r^8 -
r_1^4 r_2^4)}} Thus the terms which depend on the individual brane
locations cancel neatly and one gets the result
\eqn\deight{\eqalign{S_{eff}^{sugra} = \kappa\int d^6Z~& ({R \over
r})^8 ~{1\over |Z-Z_1|^4}{1\over |Z_2-Z|^4}\cr &\int d^4p~32 ({r \over
R})^8~[(p_a(\fa)^{ab})(p^c(\fb)_{cb}) +
(p_a(\fat)^{ab})(p^c(\fbt)_{cb})]}} The powers of ${R \over r}$ also
cancel leaving with a $Z$ integral which is \eqn\dnine{\int
d^6Z~{1\over |Z-Z_1|^4}{1\over |Z_2-Z|^4} \sim {1\over |Z_1 - Z_2|^2}}
Transforming back into (brane) position space we get \dten.

In flat space $a_1=a_2=1$ and $b_1=b_2=0$ while the factors of 
$({R \over r})^8$ are not present in the action and in \dfive. Using \dfive\
and \dsix\ it is easy to see that in this case too we reproduce exactly
\deight.
\listrefs
\end